\documentclass[preprint,12pt,authoryear]{elsarticle}

\usepackage{natbib}
\usepackage{epsfig}
\usepackage{graphicx}
\usepackage{epstopdf}
\usepackage{amsmath}
\usepackage{amssymb}

\journal{Physics Letters A}

\begin{document}

\begin{frontmatter}



\title{Research on phases of Grover-type algorithms}

\ead{2010thzz@sina.com}
\author[label1,label2]{Bo-wen Ma}
\author[label1,label2]{Wan-su Bao\corref{mycorrespondingauthor}}
\author[label1,label2]{Xiang Wang}
\author[label1,label2]{Xiang-qun Fu}
\author[label1,label2]{Shuo Zhang}
\author[label1,label2]{Tan Li}
\author[label1,label2]{Feng-guang Li}
\cortext[mycorrespondingauthor]{ correspondingauthor }
\address[label1]{Zheng-zhou Information Science and Technology Institute,Zheng-zhou 450004,China}
\address[label2]{Synergetic Innovation Center of Quantum Information and Quantum Physics, University of Science and Technology of China,Hefei,Anhui 230026,China}

\begin{abstract}
When applying Grover's algorithm to an unordered database, the probability of obtaining correct results usually decreases as the quantity of target increases. To amend the limitation, numbers of improved schemes are proposed. In this paper, we focus on four improved schemes from phases, and find that they are just differed by a global phase. Based on this conclusion, the extensive researches on one scheme can be easily generated to other three schemes, and some examples are presented to indicate the correctness.
\end{abstract}

\begin{keyword}
Grover algorithm,  global phase,  application


\end{keyword}

\end{frontmatter}


\section{Introduction}
\label{intro}

Quantum search algorithm [1-5] can be used search a target in a parallel way, compared with classical search algorithms, it achieves quadratic acceleration when searching a target in an unordered database. However, due to the property of quantum mechanics, it cannot work out an answer with certainty but with a probability. Grover's algorithm [1] is one of the most famous quantum search algorithms, nevertheless, there are still some imperfections with it. When the proportion of target is over 1/4, the success probability decreases rapidly, and when the proportion of target is over 1/2, the algorithm fails.

To amend these deficiencies, many methods have been proposed [6-14] from initial states, Hadamard-transform and phase factors. Biron D [8] and Biham E [9] extend the initial state from the superposition state to arbitrary state, demonstrated the feasibility that arbitrary state can be used to be the initial state of Grover's algorithm. Grover [12] and Tulsi [12] proposed the Hadamard-transform in original Grover algorithm can be replaced by any chosen transform. From phase factors, there are also numbers of schemes have been presented to generate Grover's algorithm, and this paper is mainly focus on phase factors. Grover [6] and Long [7] extended the phases of the algorithm from $\pi$ to arbitrary. Li D F et al. [15] proposed a new method to generate Long's scheme, and the choice of phases is more extensive. In Ref.[16], Li C M et al. generated the algorithm to four phases and put up with a more widely-used phase-matching condition. In 2007, Li P C et al. [17] proposed a three-phase algorithm and a new phase-matching condition, we called the four algorithms as Grover-type algorithms. Based on these four Grover-type algorithms, lots of extensive researches were proposed. F.M.Toyama [18] put up with a multi-phase matching subject based on Li P C's algorithm, he showed that a success probability between 99.8\% and 100\% can be yielded for the proportion of target equals to 1/10 or larger with six iterations. On the basis of Long's algorithm, Zhong et al. [19] obtained a quantum search algorithm with the success probability larger than 93.43\% with the phase 1.018, Li T et al. [20] proposed his quantum search algorithm based on multi-phase, of which success probability rises with the increases of the number of phases with just one iteration, and tends to be 100\% when the proportion of target is over a limit. In 2017, Guo Y et al. [21] proposed a Q-learning-based adjustable fixed-phase quantum Grover search algorithm, it avoids the optimal local situations, enabling success probabilities to approach one.

In this paper, we mainly focus on phase factors in four Grover-type algorithms, and a phase-transform condition is also proposed. With this phase-transform condition, if the initial states are the same, four Grover-type algorithms can be transformed to each other. When applying the four Grover-type algorithms to search the same unordered database, after the transform, the success searching probabilities of the four algorithms are identical even though the amplitudes are not same, so they can be defined to be equivalent. Based on this conclusion, many extensive researches from one scheme can be easily generated to other three schemes. For example, in Ref[LI P C], Li P C et al. mentioned that when the proportion of target is over $1/3$, the success probability is greater than $25/27$ with only one iteration, we will show that this conclusion can be generated to other three algorithms through our phase-transform condition.

This paper is organized as follows: In Section 2, Grover's algorithm and four improved Grover-type algorithms are introduced briefly. Section 3 is used to show that the four Grover-type algorithms can be transformed to each other through the phase-transform condition. Section 4 gives an example to show that the extensive results from one scheme can be transformed to others. Section 5 summarizes the whole letter.

\section{Original Grover's algorithm and four Grover-type algorithms}
\label{sec:1}

When searching through an N-elements searching space $\left\{ 0,1,2\cdots N-1\right\}( N={{2}^{n}} )$, these elements can be stored in n bits, and there are M targets for searching, $1\le M\le N$.

The initial state of the algorithm is the equal superposition state $\left| s \right\rangle$
\begin{equation}
\left| s \right\rangle ={{H}^{\otimes n}}{{\left| 0 \right\rangle }^{\otimes n}}=\frac{1}{{{2}^{n/2}}}\sum\limits_{x=0}^{{{2}^{n}}-1}{\left| x \right\rangle }=\frac{1}{\sqrt{N}}\sum\limits_{x=0}^{N-1}{\left| x \right\rangle}
\end{equation}
Grover's algorithm consists of repeated application of a quantum subroutine, called Grover iteration, denoted as G, which may be broken up into four steps:

1. Apply the oracle ${{I}_{t}}$. The purpose of using oracle ${{I}_{t}}$ is to reverse the amplitude of the target, which is ${{I}_{t}}\left| x \right\rangle ={({-1})^{f(x)}}\left| x \right\rangle $, when $\left| x \right\rangle =\left| t \right\rangle ,f(x)=1$, when $\left| x \right\rangle \ne \left| t \right\rangle ,f(x)=0$. $\left| t \right\rangle $ is the target state.

Therefore, the ${{I}_{t}}$ operator can be denoted as
\begin{equation}
{{I}_{t}}=I-2\left| t \right\rangle \left\langle  t \right|
\end{equation}

2. Perform the Hadamard-transform ${{H}^{\otimes n}}$.

3. Apply a conditional phase shift, which performs a $(-1)$ phase shift to all states except $\left| 0 \right\rangle $. This transform can be expressed as
\begin{equation}
{{I}_{0}}=2\left| 0 \right\rangle \left\langle  0 \right|-I
\end{equation}

4. Perform the Hadamard-transform ${{H}^{\otimes n}}$.

It is useful to note that the combined effect of steps 2, 3 and 4
\begin{equation}
{{I}_{s}}={{H}^{\otimes n}}\left( 2\left| 0 \right\rangle \left\langle  0 \right|-I \right){{H}^{\otimes n}}=2\left| s \right\rangle \left\langle  s \right|-I
\end{equation}

Thus Grover iteration may be written as
\begin{equation}
G={{I}_{s}}{{I}_{t}}
\end{equation}

In fact, Grover iteration can be seen as a rotation in the two-dimensional space spanned by the vector $\left| \alpha  \right\rangle $ and $\left| \beta  \right\rangle $. $\left| \alpha  \right\rangle $ indicates the normalized states of the sum of all targets, and $\left| \beta  \right\rangle $  indicates normalized states of the sum of non-targets. The initial state  $\left| s \right\rangle $ may be rewritten as
\begin{equation}
\left| s \right\rangle =\sin \theta \left| \alpha  \right\rangle +\cos \theta \left| \beta  \right\rangle
\end{equation}
where $\sin \theta =\sqrt{M/N}$.

Apply G to $\left| s \right\rangle $ for k times, and use some simple algebra,
\begin{equation}
{{G}^{k}}\left| s \right\rangle =\sin \left( \left( 2k+1 \right)\theta  \right)\left| \alpha  \right\rangle +\cos \left( \left( 2k+1 \right)\theta  \right)\left| \beta  \right\rangle
\end{equation}
when this occurs, a target will be searched  with the success probability
\begin{equation}
P={{\sin }^{2}}\left( \left( 2k+1 \right)\theta  \right)
\end{equation}
set [23]
\begin{equation}
k=\left\lfloor \frac{\pi \sqrt{N/M}}{4} \right\rfloor
\end{equation}
The image of P is shown in FIG. 1.
\begin{figure}
  \centering
  \includegraphics[width=3.6in,height=2.2in]{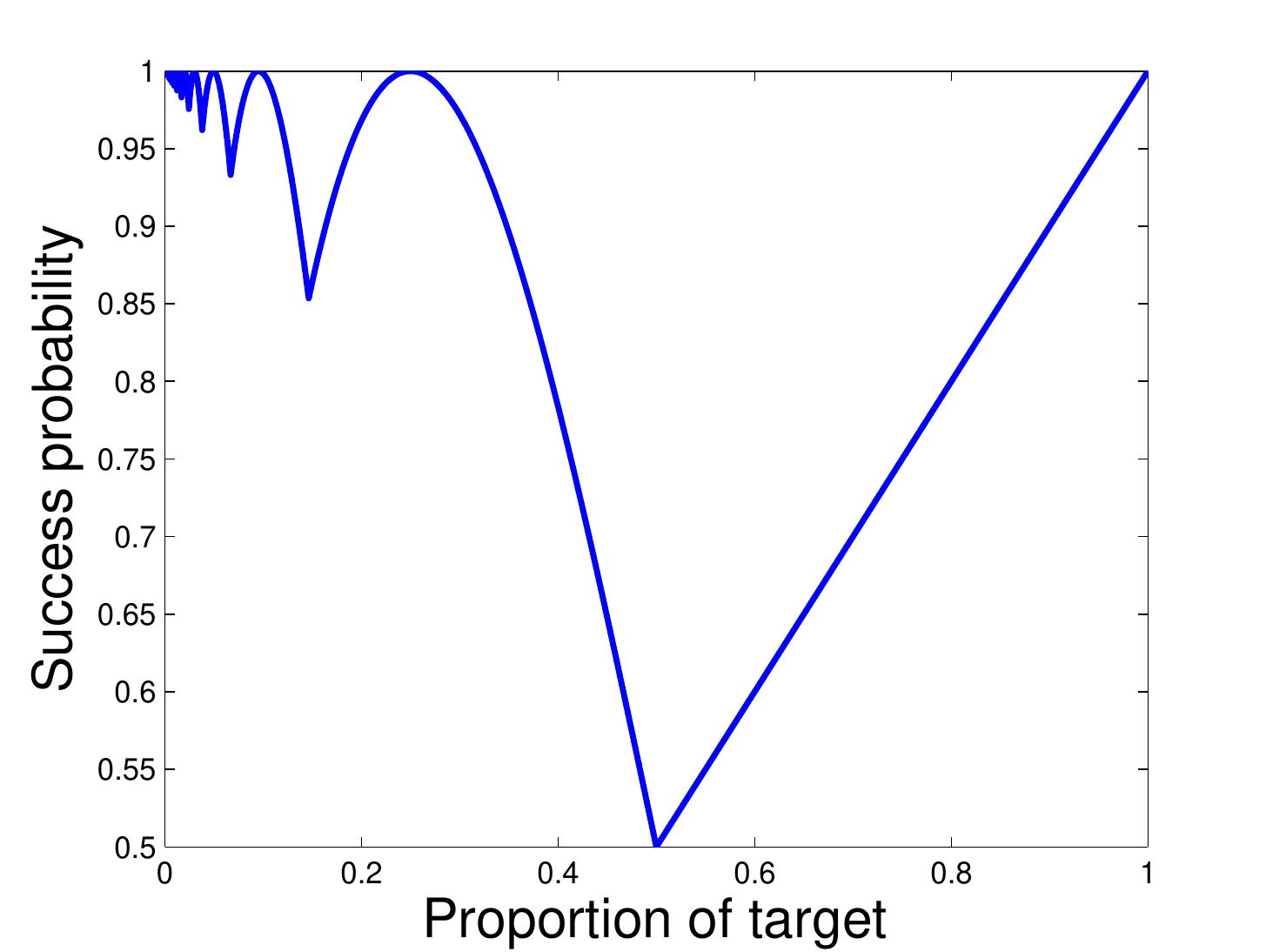}
  \caption{\textbf{ The success probability as a function of the proportion of target in Grover's algorithm. }}
  \label{1}
\end{figure}

For simplicity, the proportion of target is denoted as $\lambda \left( \lambda =M/N \right)$.
From FIG.\,1, when $1/4\le \lambda \le 1/2$, the success probability decreases rapidly, and when $\lambda \ge 1/2$, the algorithm fails. From  equation (8) and (9), when $\lambda =0.147, P=0.854$, when $\lambda =0.5, P=0.5$.

Then, four Grover-type algorithms will be introduced, all of them generate original Grover algorithm from phases.

Long's algorithm [7]
\begin{equation}
\left\{ \begin{matrix}
   {{I}_{s}}^{\left( 1 \right)}=(1-{{e}^{i\phi }})\left| \text{s} \right\rangle \left\langle  \text{s} \right|-I  \\
   {{I}_{t}}^{\left( 1 \right)}=I-(1-{{e}^{i\phi }})\left| t \right\rangle \left\langle  t \right|  \\
\end{matrix} \right.
\end{equation}

Li D F's algorithm [16]
\begin{equation}
\left\{ \begin{matrix}
   {{I}_{s}}^{\left( 2 \right)}=2\cos \tau {{e}^{i\tau }}\left| \text{s} \right\rangle \left\langle  \text{s} \right|-I  \\
   {{I}_{t}}^{\left( 2 \right)}=I-2\cos \tau {{e}^{i\tau }}\left| t \right\rangle \left\langle  t \right|  \\
\end{matrix} \right.
\end{equation}

Li C M's algorithm [17]
\begin{equation}
\left\{ \begin{matrix}
   {{I}_{s}}^{\left( 3 \right)}=({{e}^{i{{\gamma }_{1}}}}-{{e}^{i{{\gamma }_{2}}}})\left| \text{s} \right\rangle \left\langle  s \right|+{{e}^{i{{\gamma }_{2}}}}I  \\
   {{I}_{t}}^{\left( 3 \right)}=-{{e}^{i{{\gamma }_{2}}}}I-({{e}^{i{{\gamma }_{1}}}}-{{e}^{i{{\gamma }_{2}}}})\left| t \right\rangle \left\langle  t \right|  \\
\end{matrix} \right.
\end{equation}

Li P C's algorithm [18]
\begin{equation}
\left\{ \begin{matrix}
   {{I}_{s}}^{\left( 4 \right)}=(1-{{e}^{i\beta }})\left| \text{s} \right\rangle \left\langle  s \right|+{{e}^{i\beta }}I  \\
   {{I}_{t}}^{\left( 4 \right)}=I-(1-{{e}^{-i\beta }})\left| t \right\rangle \left\langle  t \right|  \\
\end{matrix} \right.
\end{equation}

For simplicity, the four algorithms mentioned above are denoted as algorithm 1, 2, 3 and 4 respectively, and the ${I}_{s}$ and ${I}_{t}$ operators are denoted as ${{I}_{s}}^{\left( i \right)}$, ${{I}_{t}}^{\left( i \right)}$, $i=1,2,3,4$.

\section{Four Grover-type algorithms are just differed by a global phase}
\label{sec:2}

\emph{Proposition 1}: When phases meets the condition $\phi =\text{2}\tau +\pi ={{\gamma }_{1}}-{{\gamma }_{2}}=-\beta $, algorithm 1, 2, 3, 4 are just differed by a global phase.

Firstly, we demonstrate on algorithm 1 \& 2.
\begin{equation}
\begin{array}{lll}
   {{I}_{s}}^{\left( 2 \right)}&=2\cos \tau {{e}^{i\tau }}\left| \text{s} \right\rangle \left\langle  \text{s} \right|-I \\
 & \text{=}2\cos \tau (\cos \tau +\sin \tau *\text{i})\left| \text{s} \right\rangle \left\langle  \text{s} \right|-I \\
 & \text{=(}2{{\cos }^{2}}\tau \text{+}2\cos \tau \sin \tau *\text{i})\left| \text{s} \right\rangle \left\langle  \text{s} \right|-I \\
 & \text{=(}\cos 2\tau +1+\sin 2\tau *i)\left| \text{s} \right\rangle \left\langle  \text{s} \right|-I \\
 & \text{=(}1+{{e}^{i2\tau }})\left| \text{s} \right\rangle \left\langle  \text{s} \right|-I \\
\end{array}
\end{equation}
set $2\tau +\pi \text{=}\phi $,
\begin{equation}
{{I}_{s}}^{\left( 2 \right)}=(1-{{e}^{i\phi }})\left| \text{s} \right\rangle \left\langle  \text{s} \right|-I
\end{equation}
which is just ${{I}_{s}}^{\left( 1 \right)}$. ${{I}_{t}}$ operator can be proved by the same methods. Then ${{G}^{\left( 1 \right)}}={{G}^{\left( 2 \right)}}$.

Next, algorithm 1 \& 3 is focused
\begin{equation}
{{I}_{s}}^{\left( 3 \right)}=({{e}^{i{{\gamma }_{1}}}}-{{e}^{i{{\gamma }_{2}}}})\left| \text{s} \right\rangle \left\langle  s \right|+{{e}^{i{{\gamma }_{2}}}}I
\end{equation}
\begin{equation}
\begin{array}{lll}
   -{{e}^{-i{{\gamma }_{2}}}}{{I}_{s}}^{\left( 3 \right)}&=-{{e}^{-i{{\gamma }_{2}}}}[({{e}^{i{{\gamma }_{1}}}}-{{e}^{i{{\gamma }_{2}}}})\left| \text{s} \right\rangle \left\langle  s \right|+{{e}^{i{{\gamma }_{2}}}}I]\\&=(-{{e}^{i({{\gamma }_{1}}-{{\gamma }_{2}})}}+1)\left| \text{s} \right\rangle \left\langle  s \right|-1 \\
\end{array}
\end{equation}
set ${{\gamma }_{1}}-{{\gamma }_{2}}=\phi $,
\begin{equation}
{{I}_{s}}^{\left( 1 \right)}=-{{e}^{-i{{\gamma }_{2}}}}{{I}_{s}}^{\left( 3 \right)}
\end{equation}
Proving by the same methods, ${{I}_{t}}^{\left( 1 \right)}=-{{e}^{-i{{\gamma }_{2}}}}{{I}_{t}}^{\left( 3 \right)}$ is obtained. Then ${{G}^{\left( 1 \right)}}={{e}^{-2i{{\gamma }_{2}}}}{{G}^{\left( 3 \right)}}$

Finally, algorithm 1 \& 4 is concentrated.

\begin{equation}
{{I}_{s}}^{\left( 4 \right)}=(1-{{e}^{i\beta }})\left| \text{s} \right\rangle \left\langle  s \right|+{{e}^{i\beta }}I
\end{equation}
\begin{equation}
\begin{array}{lll}
   -{{e}^{-i\beta }}{{I}_{s}}^{\left( 4 \right)}&=-{{e}^{-i\beta }}(1-{{e}^{i\beta }})\left| \text{s} \right\rangle \left\langle  s \right|-1 \\
 & =(-{{e}^{-i\beta }}+1)\left| \text{s} \right\rangle \left\langle  s \right|-1 \\
\end{array}
\end{equation}
set $-\beta =\phi $, then
\begin{equation}
-{{e}^{-i\beta }}{{I}_{s}}^{\left( 4 \right)}={{I}_{s}}^{\left( 1 \right)}
\end{equation}
Since ${{I}_{t}}^{\left( 4 \right)}={{I}_{t}}^{\left( 1 \right)}$, then ${{G}^{\left( 1 \right)}}=-{{e}^{-i\beta }}{{G}^{\left( 4 \right)}}$.

From the above, when phases meets the condition $\phi =\text{2}\tau +\pi ={{\gamma }_{1}}-{{\gamma }_{2}}=-\beta $, algorithm 1, 2, 3, 4 are just differed by a global phase.

\section{Applications}
\label{sec:3}

In Ref[18], Li P C et al. has proved that when the proportion of target is over $1/3$, set the phase $\beta =-\pi /2$, the success probability is greater than $25/27$ with only one iteration. We will show that this proposition can be easily generated to other three algorithms with the phase-condition, and the correctness will be proved.

From the phase-condition, when $\phi =-\beta $, the only difference between Long and Li P C's algorithm is a global phase, due to the global phase can be ignored when considering the probability, so we will draw such a conclusion, that is when $\phi =-\beta =\pi /2$, when the proportion of target is over $1/3$, the success probability is greater than $25/27$ with only one iteration in Long's algorithm. Then the correctness will be proved.

Apply Long's Grover iteration to the initial state $\left| \text{s} \right\rangle $, we will get the state $\left| {{\psi }_{1}} \right\rangle $
\begin{equation}
\begin{array}{lll}
 \left| {{\psi }_{1}} \right\rangle &={G}'\left| s  \right\rangle  \\
 &  =\left[ \left( 1-{{e}^{i\phi }} \right)\left| s \right\rangle \left\langle  s \right|-I \right]\left[ I-\left( 1-{{e}^{i\phi }} \right)\left| t \right\rangle \left\langle  t \right| \right]\\
 &*\left( \sin \theta \left| \alpha  \right\rangle +\cos \theta \left| \beta  \right\rangle  \right) \\
 &  ={{a}_{1}}|\alpha \rangle +{{b}_{1}}|\beta \rangle  \\
\end{array}
\end{equation}
where
\begin{equation}
{{a}_{1}}=\sin \theta \left[ 1-2{{e}^{i\phi }}-{{\left( 1-{{e}^{i\phi }} \right)}^{2}}{{\sin }^{2}}\theta  \right]
\end{equation}
measure $\left| {{\psi }_{1}} \right\rangle $, a target will be obtained by the probability $P={{\left| {{a}_{1}} \right|}^{2}}$. Set $\phi \text{=}\frac{\pi }{2}$ and denote $m={{\sin }^{2}}\theta $,


\begin{equation}
P=4{{m}^{3}}-8{{m}^{2}}+5m
\end{equation}
that is just the equation(14) in Ref[Li P C], so the same conclusion can be obtained, that is when the proportion of target is over $1/3$, the success probability is greater than $25/27$ with only one iteration.

Then Li D F's algorithm is concentrated. From the phase-condition, when $\tau =-\pi /4$, the proposition can be generated to Li D F's algorithm.

Applying the Grover iteration of Li D F's algorithm to the initial state $\left| \text{s} \right\rangle $,
\begin{equation}
\begin{array}{lll}
   \left| {{\psi }_{1}}' \right\rangle &=G\left| s  \right\rangle  \\
 & =\left[ 2\cos \tau {{e}^{i\tau }}\left| s \right\rangle \left\langle  s \right|-I \right]\left[ 2\cos \tau {{e}^{i\tau }}\left| t \right\rangle \left\langle  t \right| \right]\left| s \right\rangle  \\
& ={{a}_{1}}'|\alpha \rangle +{{b}_{1}}'|\beta \rangle  \\
\end{array}
\end{equation}
set $\tau =-\pi /4$, the success probability $P={{\left| {{a}_{1}}' \right|}^{2}}$ can be simplified as
\begin{equation}
P=4{{m}^{3}}-8{{m}^{2}}+5m
\end{equation}
where $m={{\sin }^{2}}\theta $. That is just the equation(14) in Ref[Li P C].

Next, L C M's algorithm is researched. From the phase-condition, when ${{\gamma }_{1}}-{{\gamma }_{2}}=\pi /2$, the proposition can be generated to Li D F's algorithm.
Apply Li C M's Grover iteration to the initial state $\left| \text{s} \right\rangle $, then
\begin{equation}
\begin{array}{lll}
  \left| {{\psi }_{1}} \right\rangle &={G}'\left| \psi  \right\rangle  \\
 & =\left[ \left( {{e}^{i{{\gamma }_{1}}}}-{{e}^{i{{\gamma }_{2}}}} \right)\left| s \right\rangle \left\langle  s \right| \text{+}{{e}^{i{{\gamma }_{2}}}} \right] \\
 & *\left[ -{{e}^{i{{\eta }_{2}}}}I-\left( {{e}^{i{{\eta }_{1}}}}-{{e}^{i{{\eta }_{2}}}} \right)\left| t \right\rangle \left\langle  t \right| \right]\left| \psi  \right\rangle  \\
 &  =-{{e}^{i{{\gamma }_{2}}}}\left[ \left( -{{e}^{i\left( {{\gamma }_{1}}-{{\gamma }_{2}} \right)}}+1 \right)\left| s \right\rangle \left\langle  s \right|-I \right] \\
 &*\left( -{{e}^{i{{\eta }_{2}}}} \right)\left[ I-\left( -{{e}^{i\left( {{\eta }_{1}}-{{\eta }_{2}} \right)}}+1 \right)\left| t \right\rangle \left\langle  t \right| \right]\left| \psi  \right\rangle \\
 &  ={{e}^{i\left( {{\gamma }_{2}}\text{+}{{\eta }_{2}} \right)}}\left[ \left( -{{e}^{i\left( {{\gamma }_{1}}-{{\gamma }_{2}} \right)}}+1 \right)\left| s \right\rangle \left\langle  s \right|-I \right] \\
 &*\left[ I-\left( -{{e}^{i\left( {{\eta }_{1}}-{{\eta }_{2}} \right)}}+1 \right)\left| t \right\rangle \left\langle  t \right| \right]\left( \sin \theta \left| \alpha  \right\rangle +\cos \theta \left| \beta  \right\rangle  \right) \\
 & \text{       =}{{a}_{1}}|\alpha \rangle +{{b}_{1}}|\beta \rangle  \\
\end{array}
\end{equation}
where
\begin{equation}
\begin{array}{lll}
& a={{e}^{i\left( {{\gamma }_{2}}\text{+}{{\eta }_{2}} \right)}}\sin \theta \left[ \left( 1-{{e}^{i\left( {{\eta }_{1}}-{{\eta }_{2}} \right)}} \right)-{{e}^{i\left( {{\gamma }_{1}}-{{\gamma }_{2}} \right)}} \right. \\
 & \text{     }\left. -\left( 1-{{e}^{i\left( {{\gamma }_{1}}-{{\gamma }_{2}} \right)}} \right)\left( 1-{{e}^{i\left( {{\eta }_{1}}-{{\eta }_{2}} \right)}} \right){{\sin }^{2}}\theta  \right] \\
\end{array}
\end{equation}
when ${{\gamma }_{1}}-{{\gamma }_{2}}\text{=}{{\eta }_{1}}-{{\eta }_{2}}=\pi /2$, the success probability $P={{\left| {{a}_{1}}' \right|}^{2}}$ can be simplified as
\begin{equation}
P=4{{m}^{3}}-8{{m}^{2}}+5m
\end{equation}
where $m={{\sin }^{2}}\theta $. That is just the equation(14) in Ref[Li P C], so the same conclusion can be obtained, that is when the proportion of target is over $1/3$, the success probability is greater than $25/27$ with only one iteration.

\emph{Example 2}

With the basis vectors $\left| \alpha  \right\rangle $ and $\left| \beta  \right\rangle $, the Grover iteration of algorithm 1 can be expressed as

\begin{equation}
{{G}^{\left( 1 \right)}}=\left[ \begin{matrix}
   -{{e}^{i\varphi }}\left( {{\sin }^{2}}\theta {{e}^{i\phi }}+{{\cos }^{2}}\theta  \right) & \sin \theta \cos \theta \left( 1-{{e}^{i\varphi }} \right)  \\
   \sin \theta \cos \theta {{e}^{i\varphi }}\left( 1-{{e}^{i\phi }} \right) & -\left( {{\cos }^{2}}\theta {{e}^{i\varphi }}+{{\sin }^{2}}\theta  \right)  \\
\end{matrix} \right]
\end{equation}
and the initial state $\left| \text{s} \right\rangle $ can be written as ${{\left( \sin \theta ,\cos \theta  \right)}^{T}}$.
Apply ${{G}^{\left( 1 \right)}}$ to $\left| \text{s} \right\rangle $ for k times, the state $\left| {{\psi }_{k}} \right\rangle $ will be obtained
\begin{equation}
\begin{array}{lll}
 |{{\psi }_{k}}\rangle &={{G}^{^{k}}}(\sin \theta |\alpha \rangle +\cos |\beta \rangle ) \\
&={{\left( {{a}_{k}},{{b}_{k}} \right)}^{T}}& \\
\end{array}
\end{equation}
measure the state $\left| {{\psi }_{k}} \right\rangle $, a target item will be searched with the probability $P={{\left| {{a}_{k}} \right|}^{2}}$.
Set k=5, and with the phase-matching condition $\phi =\varphi $, the relationship among P, phase $\phi $ and the proportion of target $\lambda$ is shown in FIG.\,2.
\begin{figure}
  \centering
  \includegraphics[width=3in,height=2in]{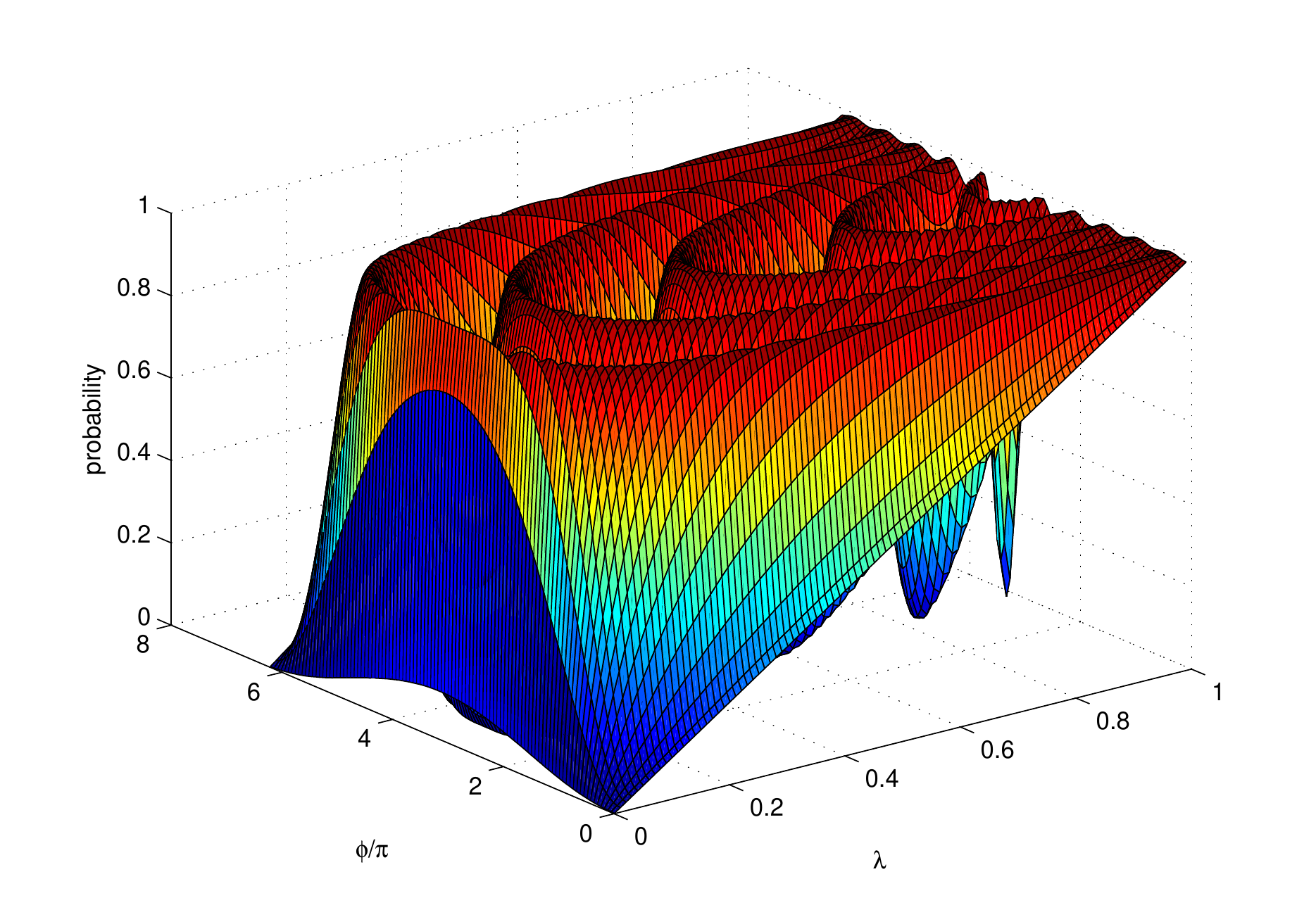}
  \caption{\textbf{  When k=5, the success probability as a function of phase and proportion of target in  algorithm 1.} }
  \label{2}
\end{figure}

Using the same method, other three images are shown in FIG.\,3-5.
\begin{figure}
  \centering
  \includegraphics[width=3in,height=2in]{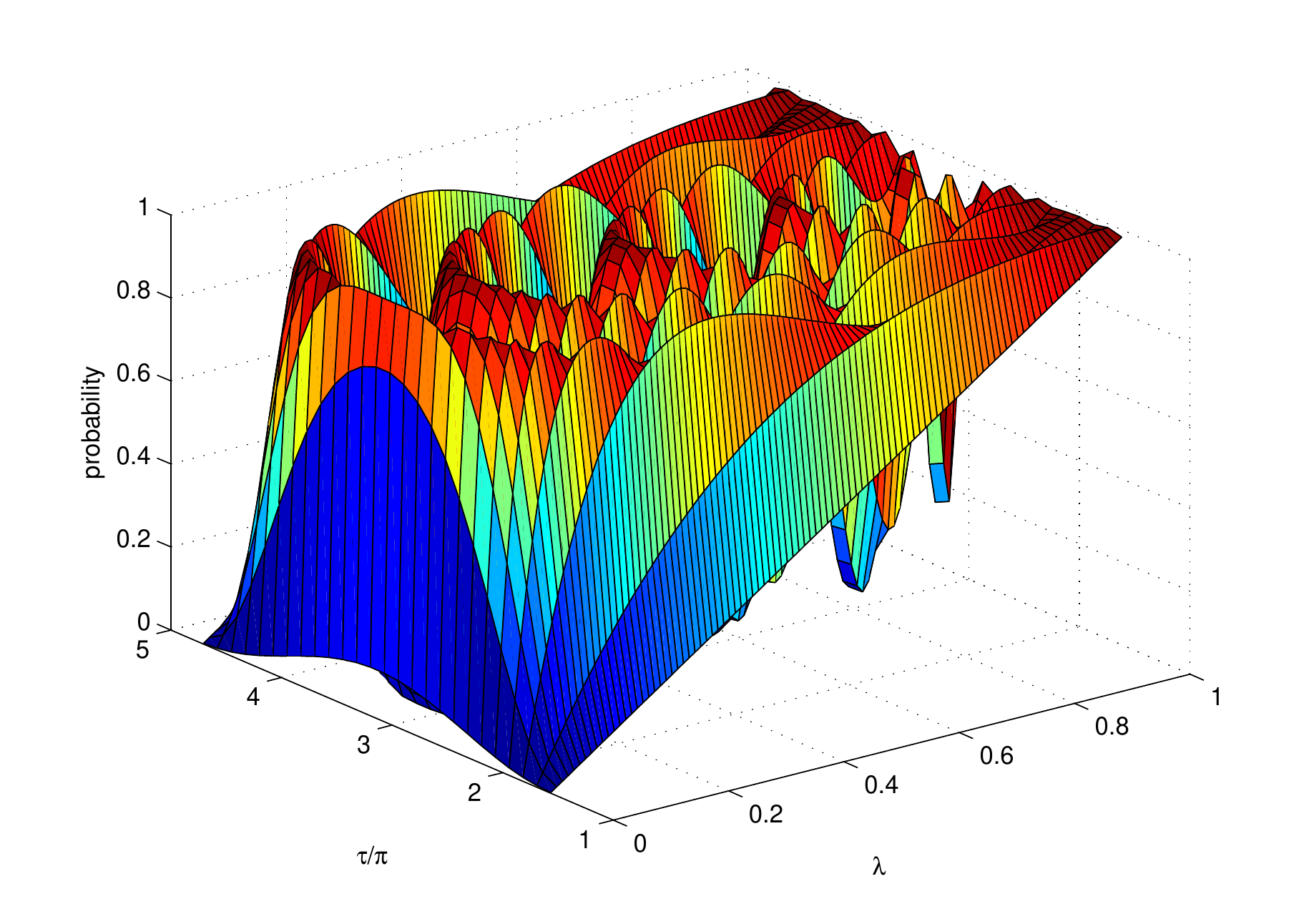}
  \caption{ \textbf{ When k=5, the success probability as a function of phase and proportion of target in  algorithm 2. }}
  \label{3}
\end{figure}

\begin{figure}
  \centering
  \includegraphics[width=3in,height=2in]{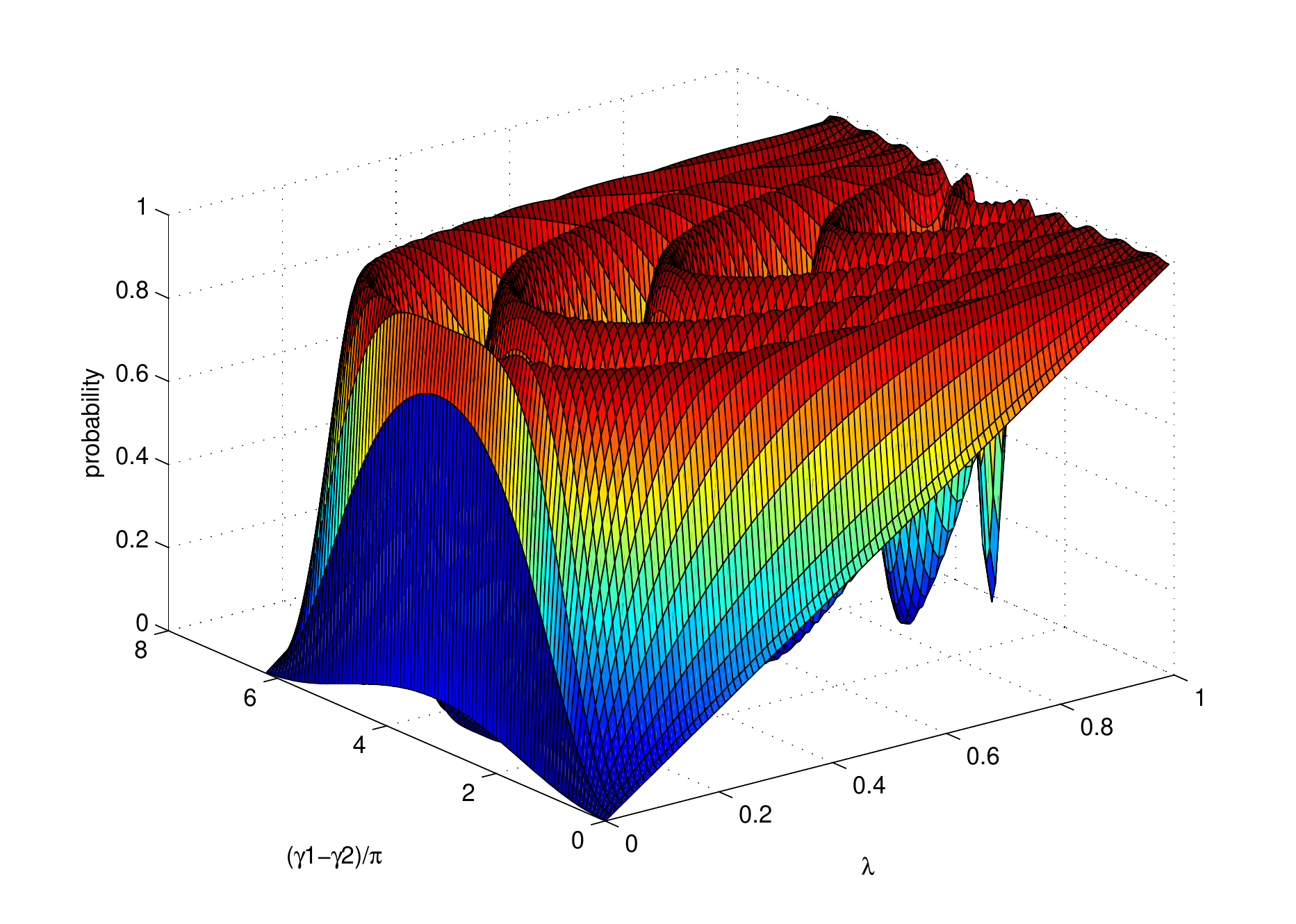}
  \caption{ \textbf{ When k=5, the success probability as a function of phase and proportion of target in  algorithm 3. }}
  \label{4}
\end{figure}

\begin{figure}
  \centering
  \includegraphics[width=3in,height=2in]{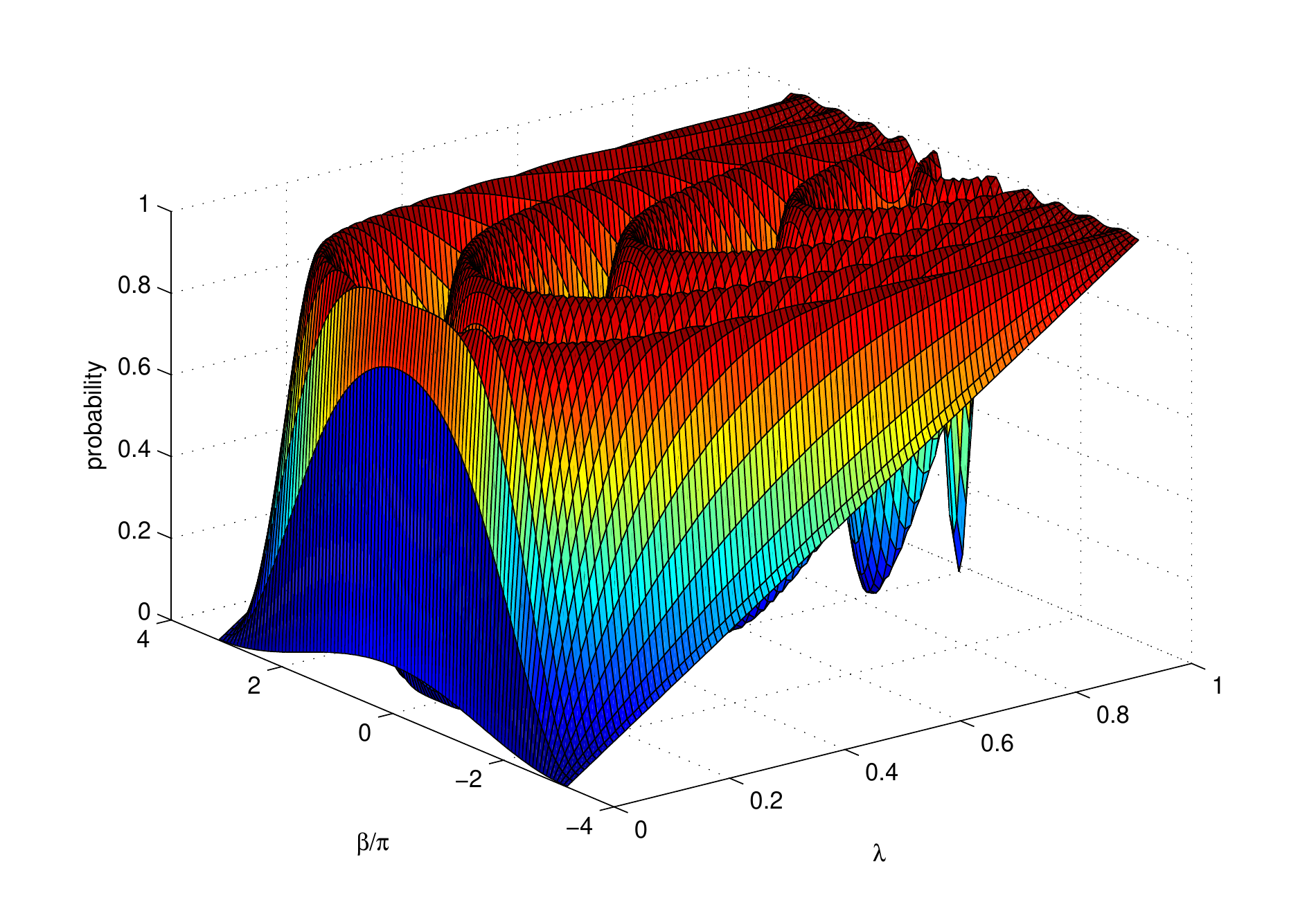}
  \caption{ \textbf{ When k=5, the success probability as a function of phase and proportion of target in  algorithm 4. }}
  \label{5}
\end{figure}

From FIG.\,2-5, when phases of algorithm 1, 2, 3 and 4 meet the condition $\phi =\text{2}\tau +\pi ={{\gamma }_{1}}-{{\gamma }_{2}}=-\beta $, the relationships among success probability, phase, and proportion of target are completely identical.

\section{Conclusion}
\label{sec:4}

In this paper, the definition of the equivalence of quantum search algorithms is proposed for the first time, and four improved schemes of Grover's algorithm are demonstrated to be equivalent. If phases of them are satisfied with a certain condition, the only difference of them is global phase, then the probability of them are identical for the same number of iteration when searching a target in the same database. Finally, the result of simulation indicates the correctness of the conclusion.

\textbf{Acknowledgments}  This work was supported by the Natural Science Foundation of China (NSFC) under Grant Nos.61502526.




\end{document}